\documentclass[10pt]{article}
\usepackage[dvips]{graphicx}

\usepackage{epsfig}
\usepackage{graphicx}
\usepackage{indentfirst}
\usepackage{amsmath}
\usepackage{amsfonts}
\usepackage{amssymb}

\begin{document}

\begin{center}
{\Large\bf Conformal anomaly around the sudden singularity\\}
\medskip

 S.J.M. Houndjo\footnote{e-mail:
sthoundjo@yahoo.fr}
\medskip

Departamento de F\'{\i}sica, Universidade Federal do Esp\'{\i}rito
Santo\\ 29060-900, Vit\'oria, Esp\'{\i}rito Santo, Brazil
\medskip

\end{center}

\begin{abstract}
Quantum effects due to particle creation on a classical sudden singularity have been investigated in a previous work. The conclusion was that quantum effects do not lead to the avoidance nor the modification of the sudden future singularity. In this paper, we investigate quantum corrections coming from conformal anomaly near the sudden future singularity. We conclude that when the equation of state is chosen to be $p=-\rho-A\rho^\alpha$, the conformal anomaly can transform the sudden singularity in the singularity of type III for any $\alpha> 1/2$ and  in the singularity of the type I (the big rip) or the big crunch for $1/2<\alpha<3/2$.
\end{abstract}

Pacs numbers: 

\section{Introduction}
There are strong evidences that the universe is in an accelereted expansion. The simplest setup to describe this acceleration is the use of a cosmological constant, with the equation  of state $p_\Lambda=-\rho_\Lambda$, where $p_\Lambda$ is the pressure and $\rho_\Lambda$  the energy density of such a cosmological constant. However, there are others candidates that astronomical datas allow. Namely, a dark energy component whose equation of state parameter, $\omega$, is very close to $-1$, where $\omega$ is the ratio between the pressure $p$ and the energy density $\rho$ of dark energy. The relevent of this point is that dark energy can lead to quite different scenarios concerning the futur of the universe. Precisely, if $\omega> -1$, dark energy corresponds to quintessence fluid; if $\omega=-1$, it is a cosmological constant and the universe would be asymptotically de Sitter;  if $\omega<-1$, dark energy corresponds to a "phantom" content which induces the occurence of singularities. Generally, the occurence of these singularities is linked with the violation of some energy conditions. For example, the known singularity which occur in the phantom case is the big rip and it violates the strong energy condition and the weak one. For this type of singularity the Hubble parameter and its cosmic time derivative diverge \cite{staro}.\par 
There exists another singularity known as sudden singularity, for which the strong energy condition is satisfied but the dominant energy condition is violated. For this singularity, the Hubble parameter is finite but its cosmic time derivative diverges \cite{barrow}. Besides to the big rip and the sudden singularity, there is others types of singularities. We present here the classification of these singularities as follow \cite{nojiri}:\newpage
$\bullet$ Type I (Big Rip): for $t\rightarrow t_s$, $a\rightarrow\infty$, $ \rho\rightarrow\infty$ and $|p|\rightarrow \infty$ or $\rho$ and $p$ are finite at $t=t_s$.\par
$\bullet$ Type II (sudden): for $t\rightarrow t_s$, $a\rightarrow a_s$, $\rho\rightarrow \rho_s$  and $|p|\rightarrow\infty$.\par
$\bullet$ Type III: for $t\rightarrow t_s$, $a\rightarrow a_s$, $\rho\rightarrow \infty$ e $|p|\rightarrow \infty$.\par
$\bullet$ Type IV: for $t\rightarrow t_s$, $a\rightarrow a_s$, $\rho\rightarrow 0$ e $p\rightarrow 0$ and higher derivatives of $H$ diverge.\par
Here, $a$ means the scale factor and $t_s$ referring the instant of the singularity. We assume the cosmic fluid to be ideal.\par
Our attention is focalized on the sudden singularity for which $a$ and $\rho$ are finite but $p$ diverges \cite{barrow}. This singularity occurs without violating the strong energy condition but violates the dominant one as mentioned above. Generally, when singularities appear, the relevent question to ask is: can the singularity be modified or eliminated by taking into account some physical effects? The answer is affirmative, due to at least two reasons. The first one is the quantum corrections due to the {\it conformal anomaly}. This is used in various cosmological applications: the dynamical Casimir efffect with conformal anomaly \cite{brevik1,brevik2}, or a dark fluid with conformal anomaly \cite{nojiri2}. The second  reason, which is not the subject of this work is to take into account the {\it bulk viscosity} of the cosmic fluid.  
A number of studies has been carried out about quantum effects near the sudden singularity and various answers have been found. In \cite{article2}, Barrow et al have shown that quantum effects due to particle creation can not prevent nor modify the future sudden singularity. In that case the quantum effects have been studied comparing the behaviour of the energy density of the particle creation with the energy density of the fluid which characterizes the classical background. The effects of loop quantum gravity have been investigated in cosmologies exhibiting classical sudden singularities and they can remove the sudden singularity under certain particular conditions \cite{sami},  and there is a close relationship between sudden singularities and the behaviour of Friedmann universes containing bulk viscous stresses \cite{barrow2}. Brevik et al \cite{brevik3} have shown that the conformal anomaly corrections causes the exponents of the future singularity to be modified. They dealt with the singularities of type I and type III. The general structure of the singularities has been studied by Nojiri \cite{nojiri} and the energy conditions about each of them presented. It is important to note that many results have been found concerning quantum effects near the big rip. In an early work \cite{article1}, it has been  shown that quantum effects due to particle creation can not prevent the big rip. It has been used the transplankian problem for calculating the energy density of particle creation and found that this energy density tends to zero when the big rip is approched. The same idea, for analysing the quantum effects near the big rip has been estudied in \cite{article3} where it has been used the $n$-wave regularization for calculating the energy density of particle creation and found that, in this case, it tends to infinity when the big rip is approched and becomes the dominant component of the universe. This means that the big rip can be avoided by a scalar massless field. The opposite of the latter conclusion has been done by Pavlov \cite{pavlov} where he dealt with a massive scalar field, but this, in a very particular case.\par
In this paper we study the quantum effects due to conformal anomaly on the sudden singularity. Specifically, we seek to know if quantum effects can modify or eliminate the classical sudden singularity. We use the conformal anomaly and investigate its influence on the classical background density. We find that, depending on the value of the parameter $\alpha$, quantum effects due to conformal anomaly can transform the sudden singularity in the singularity of type I (big rip), or in the singularity of the type III or also in the big crunch.\par 
The paper is organized as follows. In the next section, we will set out the sudden singularity models. In section 3 the conformal anomaly
near the singularities in general and in particular its effects near the sudden singularity will be presented. The analyse will be done in the case for which the thermodynamical parameter $\omega$ in the equation of state $p=\omega \rho$, depends on the time . Note that it is impossible to have a constant thermodynamic parameter in the sudden singularity, since the pressure is singular while the density is finite. In the section 4 we present the conclusions.

\section{Sudden singularity model}
We considere the most general form of the sudden singularity which can occur at finite time  in a Friedmann expanding universe as discussed in \cite{barrow}. It has been shown that  singularities of this sort can arise from a divergence of the pressure, $p$, at finite time despite the scale factor, $a(t)$, the material density, $\rho$, and the Hubble parameter rate, $H=\dot{a}/a$, remainning finite. The pressure singularity is connected with the divergence in the acceleration of the universe, $\ddot{a}$, at finite time. This singularity occurs without violating the strong energy condition  $\rho+3p>0$ which is not the case of the big rip singularity. The only way to prevent the sudden singularity is to bound the pressure.\par
Considere the spatially flat isotropic Friedmann universe. The Einstein equation leads to
\begin{eqnarray}
3\biggr(\frac{\dot{a}}{a}\biggl)^{2} &=& 8\pi G\rho ,  \label{1} \\
\frac{\ddot{a}}{a} &=&-\frac{4\pi G}{3}(\rho +3p),  \label{2} \\
\dot{\rho}+3\frac{\dot{a}}{a}(\rho +p) &=&0.  \label{3}
\end{eqnarray}
In \cite{barrow}, it has been constructed an explicit example for seeking, over the time interval $0<t<t_s$, a suitable solution for the scale fator $a(t)$,  of the form
\begin{eqnarray}
a =\biggr(\frac{t}{t_{s}}\biggl)^{q}(a_{s}-1)+1-\biggr(1-\frac{t}{t_{s}}%
\biggl)^{n}.  \label{a}
\end{eqnarray}
Hence, the first and second derivatives of the scale factor remain
\begin{eqnarray}
\dot{a} &=&\frac{q}{t_{s}}\biggr(\frac{t}{t_{s}}\biggl)^{q-1}(a_{s}-1)+%
\frac{n}{t_{s}}\biggr(1-\frac{t}{t_{s}}\biggl)^{n-1},  \label{b} \\
\ddot{a} &=&\frac{q(q-1)}{t_{s}^{2}}\biggr(\frac{t}{t_{s}}\biggl)%
^{q-2}(a_{s}-1)-\frac{n(n-1)}{t_{s}^{2}}\biggr(1-\frac{t}{t_{s}}\biggl)%
^{n-2}.
\end{eqnarray}%
In order to have the divergence in the second derivative while the scale factor and his first derivative remain finite, one imposes $1<n<2$ and $0<q\leq1$. As $t\rightarrow t_s$, we have $a\rightarrow a_s$, $H\rightarrow H_s$ and $\rho\rightarrow\rho_s$ where $a_s$, $H_s$ and $\rho_s$ are all finite but $p\rightarrow\infty$. As $t\rightarrow t_s$, due to the fact that the density is finite and the pressure divergent, the dominant energy condition, $|p|\leq\rho$, must always be violated. \\
The equation of state in the sudden singularity case, which we will choose to be the same even when conformal anomaly is taken into account, is written as 
\begin{equation}\label{state}
p=-\rho-f(\rho) .
\end{equation}
The divergence which occurs in the pressure $p$ must be incorporated by $f(\rho)$. Since the energy density $\rho$ is finite near the sudden singularity, we choose $f(\rho)$ as
\begin{eqnarray}
f(\rho)&=&A\left(\rho_s-\rho\right)^{-\lambda},\label{fro}\\
\rho&=&\rho_s+\rho_0\left(t_s-t\right)^{\sigma}\label{ro},
\end{eqnarray}
where $\lambda$ and $\sigma$ are positive constants, $\rho_s$ the energy density at the singularity time and $\rho_0$ a positive constant. This expression of the energy density is used due to the fact that the form of the scale fator in (\ref{a}) does not let tractable the expression of the energy density through the equation (\ref{1}). It is clear that, as $t\rightarrow t_s$, the energy density remains finite while the pressure diverges.

\section{Conformal anomaly near the sudden singularity}
The study of the quantum effects near the singularities is of main interest. Here, we will study the effect of quantum backreaction of conformal matter around the sudden singularity. The conformal anomaly, also known as trace anomaly, is the fact that the trace of the energy momentum tensor (EMT) vanishes for the classical conformally coupled field, $g^{\mu\nu}T_{\mu\nu}=0$, while the trace of the renormalized expectation value , $g^{\mu\nu}\left\langle T_{\mu\nu}\right\rangle_{ren}$, is non zero. \par
For any classical field with conformally invariant action, e.g. for conformally coupled massless scalar fields or for the electromagnetic field, the trace of the EMT identically vanishes. This can be shown by a simple calculation. If the action $S[\phi, g_{\mu\nu}]$ of a general covariant theory is invariant under conformal transformations, 
\begin{eqnarray}
S[\phi,g_{\mu\nu}]= S[\phi,\bar{g}_{\mu\nu}], \quad\quad g_{\mu\nu}(x)\longrightarrow \bar{g}_{\mu\nu}(x)=\Omega^{2}(x)g_{\mu\nu}(x),
\end{eqnarray}
where $\Omega(x)\neq 0$ is an arbitrary smooth function, then the variation of the action with respect to an infinitesimal conformal transformation must vanish. An infinitesimal conformal transformation with  $\Omega(x)=1+\delta\Omega(x)$ yields
\begin{eqnarray}
\delta g_{\mu\nu}(x)=2g_{\mu\nu}(x)\delta\Omega(x), \quad\quad \delta g^{\mu\nu}(x)=-2g^{\mu\nu}(x)\delta\Omega(x).
\end{eqnarray}
Using the definition 
\begin{eqnarray}\label{tracedef}
T_{\mu\nu}\equiv \frac{2}{\sqrt{-g}}\frac{\delta S_m}{\delta g^{\mu\nu}} ,
\end{eqnarray}
where $S_m$ is the action of the matter field, we find
\begin{eqnarray}\label{zeroaction}
0=\delta S_m=\int d^4x\frac{\delta S}{\delta g^{\mu\nu}(x)}\delta g^{\mu\nu}(x)=\int d^4x\sqrt{-g}T_{\mu\nu}g^{\mu\nu}\delta\Omega.
\end{eqnarray}
This relation should hold for arbitrary functions $\delta \Omega(x)$, therefore the integrand must vanish for all $x$,
\begin{eqnarray}\label{nulltrace}
T^{\mu}_{\mu}(x)\equiv T_{\mu\nu}g^{\mu\nu}(x)\equiv 0.
\end{eqnarray} 
This conclusion holds for any classical generally covariant and conformally  invariant field theory, but fails for quantum fields. This failing is due to the fact that when one compute EMT in some quantum vacuum state, divergencies appear and it necessary to renormalize the quantum EMT. In a renormalization process, the divergence parts wich appear in the expression of the quantum energy momentum tensor is discarded. Consequently, the trace of the energy momentum tensor turns out to be non null. Various tecnics of regularization can be used for renormalizing the quantum energy momentum tensor and specifically, for a massless conformal scalar field in a conformally invariant space-time, the renormalized energy momentum tensor is \cite{bunch1,deser,christensen,duff,tsao}
\begin{eqnarray}
\left\langle T_{\mu\nu}\right\rangle_{ren}&=&-\frac{1}{2880\pi^2}\Bigg[\left(-\frac{1}{3}R_{;\mu\nu}+R^{\,\,\alpha}_{\mu}R_{\alpha\nu}-RR_{\mu\nu}\right)\nonumber\\
&+&g_{\mu\nu}\left(\frac{1}{3}\Box R-\frac{1}{2}R^{\alpha\beta}R_{\alpha\beta}+\frac{1}{3}R^2\right)\bigg],
\end{eqnarray}
whose the trace is
\begin{eqnarray}
g^{\mu\nu}\left\langle T_{\mu\nu}\right\rangle_{ren}&=&-\frac{1}{2880\pi^2}\left(\Box R-R^{\alpha\beta}R_{\alpha\beta}+\frac{1}{3}R^2\right)\nonumber\\
&=&-\frac{1}{2880\pi^2}\left(R^{\alpha\beta\gamma\delta}R_{\alpha\beta\gamma\delta}-R^{\alpha\beta}R_{\alpha\beta}+\Box R\right).
\end{eqnarray} 
Since the quantum corrections usually contain the powers of the curvature or higher  derivatives terms, such correction terms plays important role near the singularity. We will use here the general form of the conformal anomaly contribution as backreaction as in \cite{brevik3}.\par
The conformal anomaly $T_A$ in its general form is written as follow:
\begin{eqnarray}\label{justine}
T_A= b\left(F_1+\frac{2}{3}\Box R \right)+b^{\prime}F_2+b^{\prime\prime}\Box R \quad,
\end{eqnarray}
where $F_1$ is the square of the Weyl tensor and $F_2$ the Gauss-Bonnet invariant, which are defined as
\begin{eqnarray}
F_1&=&C^{\alpha\beta\gamma\delta}C_{\alpha\beta\gamma\delta}\nonumber\\&=&\frac{1}{3}R^2-2R^{\alpha\beta}R_{\alpha\beta}+R^{\alpha\beta\gamma\delta}R_{\alpha\beta\gamma\delta}\quad,
\end{eqnarray}

\begin{eqnarray}
F_2=R^2-4R^{\alpha\beta}R_{\alpha\beta}+R^{\alpha\beta\gamma\delta}R_{\alpha\beta\gamma\delta}.
\end{eqnarray}
In general, when we have $N$ scalars, $N_{1/2}$ spinors, $N_1$ vector fields, $N_2$ gravitons and $N_{HD}$ higher derivative conformal scalars, the coefficients $b$ and $b^{\prime}$ are given by the expressions
\begin{equation}
b=\frac{N+6N_{1/2}+12N_1+611N_2-8N_{HD}}{30(8\pi)^2}
\end{equation}

\begin{equation}
b^{\prime}=\frac{N+11N_{1/2}+62N_1+1411N_2-28N_{HD}}{10(24\pi)^2}
\end{equation}
whereas $b^{\prime\prime}$ is an artibrary constant whose value depends on the regularization method used. For the usual matter we have $b>0$ and $b^{\prime}<0$, except for higher derivative conformal scalars.\par
Quantum effects due to the conformal anomaly act as a fluid with energy density $\rho_A$ and pressure $p_A$. The total energy density is $\rho_{tot}=\rho+\rho_A$. The conformal anomaly also known as trace anomaly $T_A$ is given by
\begin{equation}\label{trace}
T_A=-\rho_A+3p_A.
\end{equation}
The density and the pressure connected with the conformal anomaly obey to the energy conservation law in the Friedmann universe as in (\ref{3}):
\begin{equation}\label{conservation}
\dot{\rho}_A+3\frac{\dot{a}}{a}(\rho_A +p_A) =0
\end{equation}
Putting (\ref{trace}) into (\ref{conservation}), one finds
\begin{eqnarray}
T_A&=&-4\rho_A-\frac{\dot{\rho}_A}{H}\nonumber\\
&=&-4\rho_A-\frac{a d\rho_A}{da},
\end{eqnarray}
which, multiplied by $a^3da$, leads to
\begin{eqnarray}
\frac{d(a^4\rho_A)}{dt}&=&-a^3\frac{da}{dt}T_A\nonumber\\
&=&-a^4HT_A.
\end{eqnarray}
Integrating, one obtains
\begin{eqnarray}
\rho_A=-\frac{1}{a^4}\int a^4 HT_Adt.
\end{eqnarray}
In terms of the Hubble parameter, the conformal anomaly (\ref{justine}) is given by
\begin{eqnarray}\label{trace2}
T_A=-12b\dot{H}^{\,2}+24b^{\,\prime}\left(-\dot{H}^{\,2}+H^{\,2}\dot{H}+H^{\,4}\right)-(4b+6b^{\,\prime\prime})\left(H^{(3)}+7H\ddot{H}+4\dot{H}^{\,2}+12H^{\,2}\dot{H}\right).
\end{eqnarray}
Taking into account the conformal anomaly, the equation (\ref{1}) turns out to be
\begin{equation}\label{friedmann2}
\frac{3}{\kappa^2}H^2=\rho+\rho_A,
\end{equation}
where $\kappa^2=8\pi G$.
Due to the fact that we are dealing with the sudden singularity, the Hubble parameter $H$ is finite, but the first and higher derivatives diverge initially before the introduction of the conformal anomaly. Also, each higher derivative of $H$ is more singular than the early one. This means that the dominant term in (\ref{trace2}) is $-(4b+6b^{\prime\prime})H^{(3)}$. In the same way, $3/(8\pi G)H^2<<|\rho_A|$, and the equation (\ref{friedmann2}) leads to $\rho \approx-\rho_A$.  Hence, we have
\begin{equation}\label{ropoint}
\dot{\rho}\approx -(4b+6b^{\prime\prime})HH^{(3)}.
\end{equation}
 When one takes quantum effects into account, the energy density of the fluid stop to be finite by a back reaction, since  $\rho_A$ is a divergente quantity and $\rho \approx-\rho_A$. In this case, we assume a new form for the state equation as 
\begin{eqnarray}\label{newse}
p&=&-\rho-f(\rho)\nonumber\\
&=&-\rho-A\rho^\alpha
\end{eqnarray}
where we impose the condition $\alpha>1/2$. This choice will be clear latter. The constant $A$ is chosen to be positive for guaranting to the pressure to be negative. Since we know now that the energy density in the presence of the conformal anomaly is divergent, one chooses the energy density in a new form as
\begin{equation}\label{newdensity}
\rho=\rho_0\left(t_s-t\right)^{-\sigma},
\end{equation}
where $\sigma$ continues being positive constant.
\par
Using the conservation law (\ref{3}) and the equation (\ref{newse}), one obtains
\begin{eqnarray}\label{hdefine}
H&=&\frac{\dot{\rho}}{3f(\rho)} \nonumber\\
&=&\frac{\dot{\rho}}{A\rho^\alpha}\quad.
\end{eqnarray}
Hence,
\begin{equation}\label{hubble}
H \propto  \left(t_s-t\right)^{-1+\sigma(\alpha-1)}.
\end{equation}
Using (\ref{hubble}), the product $H^{(3)}H$ behaves as
\begin{equation}\label{c1}
H^{(3)}H\propto \left(t_s-t\right)^{-5+2\sigma(\alpha-1)},
\end{equation}
which also is the behaviour of $\dot{\rho}$ through the equation (\ref{ropoint}). Integrating (\ref{c1}), one obtains
\begin{equation}\label{c2}
\rho\propto\left(t_s-t\right)^{-4+2\sigma(\alpha-1)}.
\end{equation}
Identifying (\ref{c2}) with (\ref{newdensity}), we obtain the parameter $\sigma$ as
\begin{equation}\label{sigma}
\sigma=\frac{4}{2\alpha-1}\quad.
\end{equation}
Since we need to have the parameter $\sigma$ as a positive quantity, it is necessary to have the condition $\alpha>1/2$ which justify the choise that we have done above.
Hence, the pressure behaves as
\begin{equation}\label{pressionfinale}
p\propto\left(t_s-t\right)^{\frac{-4\alpha}{2\alpha-1}}.
\end{equation}
With the condition $\alpha>1/2$, the quantity $\frac{-4\alpha}{2\alpha-1}$ is negative. This means that the divergence in the pressure remains. Note that, when the quantum correction becomes important, its works to furnish a negative energy density $\rho_A$ which pratically cancels with the energy density $\rho$ of the dark energy. The conformal anomaly contribution leads to a divergent energy density while the pressure continues being infinite. Although, the important question to be asked is to know what happens about the scale factor after the introduction of the conformal anomaly. There is only two possibilities: the scale factor can remain finite as in the sudden singularity case or become infinite. Thus, the quantum effects due to the conformal anomaly can transform the sudden singularity in the singularity of type III, in the singularity of type I (the big rip) or in the big crunch depending on the value of $\alpha$. Note that the singularity of type III is a  singularity for which, at a finite time, the scale fator is finite while the pressure and the energy density are divergent. Thus, it is clear that quantum effects coming from the conformal anomaly work to modify the singular nature of the universe. But when one analyses the behaviour of the Hubble parameter after the introduction of the conformal anomaly, one see that it behaves as
\begin{eqnarray}\label{hubblenovo}
H \propto \left(t_s-t\right)^{\frac{2\alpha-3}{2\alpha-1}}\quad,
\end{eqnarray}
and its rate behaves as 
\begin{eqnarray}\label{ratehubble}
\dot{H}\propto \left(t_s-t\right)^{\frac{-2}{2\alpha-1}}.
\end{eqnarray}
This allow us to find the conditions on $\alpha$ for obtainning the results coming from the conformal anomaly effects.\par
Hence, we see that for $1/2<\alpha<3/2$, the Hubble parameter turns out to be divergent and for $\alpha>3/2$ it remains finite as near the sudden singularity without conformal anomaly. But for any $\alpha>1/2$, the rate of the Hubble parameter $\dot{H}$ is divergent. When the Hubble parameter diverges, this means that we have three possibilities:\par
$\bullet$  $\dot{a}$ diverges and $a$ is constant, or \par 
$\bullet$   $\dot{a}\neq 0$ and $a\rightarrow 0$, or \par
$\bullet$ $\dot{a}$ and $a$ diverge but $\dot{a}$ diverges more than $a$.\\
In the first case, it is clear that the scale factor is finite and besides to this, the energy density and the pressure of the fluid are divergent: this is the singularity of the type III. In the second case, since the scale factor goes toward zero, the singularity is the big crunch. In the third case, the scale factor is divergent, knowing that the energy density and the pressure of the fluid are also divergent: this is the big rip. Hence, for $1/2<\alpha<3/2$, we conclude that the conformal anomaly can transform the sudden singularity in the singularity of type III or in the singularity of type I (big rip) or also in  the big crunch.\par
When the Hubble parameter is constant, we assume that the unique possibility is:\\
$\bullet$ $\dot{a}$ and $a$ are finite but $a\neq 0$.\\
Since the scale factor is finite is this case, we conclude that for $\alpha>3/2$, the conformal anomaly can transform the sudden singularity only into the singularity of type III.
\section{Conclusion}
We have evaluated the quantum effects due to the conformal anomaly  around the sudden singularity. The sudden singularity is a singularity which occurs at finite time, by the divergence appearing in the pressure while the scale factor and the energy density remain finite. This means also that the first derivative of the Hubble parameter is singular while the proper Hubble parameter remains finite. Since the conformal anomaly brings some higher derivative contribution of the Hubble parameter, this leads to a divergent energy density without eliminate the singular behaviour of the pressure and the state equation is assumed to be  $p=-\rho-A\rho^\alpha$. Thus, quantum effects coming from the conformal anomaly can modify the nature of the sudden singularity  in an expanding universe. Depending on the latter form of the scale factor (remaining finite or becoming singular) after the conformal anomaly effects, the sudden singularity can be transformed in the singularity of type III, or in the singularity of type I (the big rip) or also in the big crunch. We see that for any $\alpha>3/2$ there is always the possibility to have a constant scale factor and then the sudden singularity could be transformed by the conformal anomaly in the singularity of type III. But the only condition for the sudden singularity being transformed in the big rip or in the big crunch is $1/2<\alpha<3/2$. We conclude also that the conformal anomaly  makes the Hubble parameter to be divergent for $1/2<\alpha<3/2$, which is not the case near the sudden singularity without the conformal anomaly contribution. Moreover, we mention that the singular behaviour of the energy density comes from the negative  energy contribution of the conformal anomaly wich cancels the dark energy one.

\vspace{0.5cm}
{\bf Acknowledgement:} The author thanks CNPq (Brazil) for partial
financial support and also Prof. J. C. Fabris for his comments and discussions.

\end{document}